\documentclass[twocolumn,showpacs,preprintnumbers,amsmath,amssymb,prb]{revtex4-1}
\usepackage{graphicx}
\usepackage{dcolumn}
\usepackage{bm}
\usepackage{color}
\usepackage{physics}

\newcommand{\e}{{\bf e}}

\newcommand{\ii}{{\rm i}}

\def\gsim{\lower.35em\hbox{$\stackrel{\textstyle>}{\textstyle\sim}$}}

\begin{document}

\title{Chiral response of twisted bilayer graphene}

\author{T. Stauber$^{1}$, T. Low$^2$, and G. G\'omez-Santos$^3$, }

\affiliation{$^{1}$ Materials Science Factory,
Instituto de Ciencia de Materiales de Madrid, CSIC, E-28049 Madrid, Spain\\
$^{2}$ Department of Electrical \& Computer Engineering, University of Minnesota, Minneapolis, Minnesota 55455, USA\\
$^3$Departamento de F\'{\i}sica de la Materia Condensada, Instituto Nicol\'as Cabrera and Condensed
Matter Physics Center (IFIMAC), Universidad Aut\'onoma de Madrid, E-28049 Madrid, Spain}
\date{\today}

\begin{abstract}
We present an effective (minimal) theory for chiral two-dimensional materials. These materials possess an electro-magnetic coupling without exhibiting a topological gap. As an example, we study the response of doped twisted bilayers, unveiling unusual phenomena in the zero frequency limit. An in-plane magnetic field induces a huge paramagnetic response at the neutrality point and, upon doping, also gives rise to a substantial longitudinal Hall response. The system also accommodates nontrivial longitudinal plasmonic modes which are associated with a longitudinal magnetic moment, thus endowing them with a chiral character. Finally, we note that the optical activity can be considerably enhanced upon doping and our general approach would enable systematic exploration of 2D materials heterostructures with optical activity.
\end{abstract}

\maketitle




{\it Introduction.} Naturally occurring optically active or chiral molecules have attracted great attention and are used in many applications.\cite{Barron04,Boriskina15} These molecules often display a spinal structure leading to molecular multipole transitions, but the scattering process can most simply be modelled by an electric {\it and} magnetic dipole. Recently, chiral plasmonic metamaterials and artificial nanostructures with enhanced chiral sensing capabilities have also been designed.\cite{Tang11,Zhao17,Guerrero11,Shen13} 

Two dimensional van der Waals materials made possible the design of atomically thin chiral metamaterials with enantiomers stacking and a novel optically active material is given by twisted bilayer graphene (TBG).\cite{Kim16} It consists of two graphene layers that are rotated by an arbitrary angle with respect to each other.\cite{Li10,Schmidt10,Brihuega12,Havener14,Schmidt14,Patel15,Ray16} Its electronic structure is characterized by two Dirac cones which are separated in the Brillouin zone by the relative angle.\cite{Lopes07,Bistritzer11} The absorption is enhanced for transitions close to the van Hove singularity that is located in between the two Dirac points.\cite{Moon13,Yin16} 

Twisted bilayer graphene is a chiral material since its left- and right-handed copies are given by the relative rotation of the two layers and are linked by mirror symmetry. Due to this property, TBG displays (strong) optical activity at finite frequencies corresponding to transitions around the M-point\cite{Kim16} which can be related to the relative rotation of the chiral electrons of the two layers.\cite{Suarez17} Linearly polarized light thus experiences a Faraday rotation without breaking the TR nor rotational symmetry and strong circular dichronism has been observed that is usually only seen in the presence of a magnetic field\cite{Poumirol17} or enantiomeric structures.\cite{Kuwata05,Rogacheva06,Plum09,Zhou12}

In this Letter, we will investigate the  response of TBG,  
focusing on the terahertz and $\omega\to0$ limit, and its doping dependence. For this, we will derive an effective (minimal) model to describe the response of general chiral 2D materials. First, there must be a minimum of two layers separated by a non-zero distance $a$, because optical activity without breaking the time-reversal symmetry is a non-local property.\cite{Landau1984electrodynamics} We also assume, as was confirmed in Refs. \cite{Kim16,Suarez17}, that in-plane currents provide an adequate description of the response and current densities perpendicular to the layers are negligible. The response is then defined by the $4\times4$ matrix $\bm \sigma$ with
\begin{equation}\label{deltadiag}
\begin{bmatrix} \bm j^{(1)} \\  \bm j^{(2)}  \end{bmatrix} = \bm \sigma
\begin{bmatrix} \bm E^{(1)} \\  \bm E^{(2)}  \end{bmatrix} 
,\end{equation}
where $\bm j^{(1,2)} $ and $\bm E^{(1,2)}$   represent in-plane currents and total fields  in the plane indices ($1$ and $2$). Let us consider a rotationally invariant system for which we define the following response:
\begin{equation}\label{response}
\bm \sigma = 
\begin{bmatrix}  \sigma_0^1 \bm 1 & \sigma_1 \bm 1  -\ii\sigma_{xy} \bm \tau_y \\ \sigma_1 \bm 1+\ii\sigma_{xy} \bm \tau_y & \sigma_0^2 \bm 1 \end{bmatrix}
,\end{equation}
where $\bm 1$ and $\bm \tau_{y} $ are the $2\times2$ Pauli matrices in coordinate indices, and $\sigma_{0,1}^n(\omega) $, $\sigma_{xy}(\omega)$  are c-functions  characterizing the response, in-plane local approximation implied. 
They can be interpreted as the in-plane conductivity in layer $n=1,2$, the covalent drag conductivity as well as the Hall or chiral conductivity, respectively. In the case of twisted bilayer graphene, the conductivities 
also depend parametrically on the twist angle $\theta$, satisfying the following parity relations: $\sigma_{0,1}^n(\omega,\theta) =\sigma_{0,1}^n(\omega,-\theta)\;, \;\sigma_{xy}(\omega,\theta) =-\sigma_{xy}(\omega,-\theta)$. 

The above form is the most general response that complies with  reciprocity for time reversal, in-plane rotational invariance ($z$-axis), and a simultaneous $\pi$ rotation around (for instance) $x$-axis and the exchange of plane indices, the latter symmetry corresponding to irrelevance of incoming side choice. However, we will set $\sigma_0^1=\sigma_0^2\equiv\sigma_0$ for simplicity in what follows, see SI for the general case.\cite{SI}

{\it Effective local description.} For chiral systems, it is common to analyze the response in terms of a magnetization. By this, we can transform the general non-local description including only electric fields into an effective local description including both electric and magnetic fields through an electro-magnetic coupling. 

To make contact with this tradition, 
 the total in-plane magnetic moment (per unit surface) is written as $\bm m_{\parallel}   = a \, \bm j_m \times \bm{\hat{z}}$,
so that the contribution $2\bm j_m=\bm j^{(1)}-\bm j^{(2)}$ which distinguishes the current in each layer can be thought of as coming from a magnetic dipole density, $\bm m_{\parallel} / a $,  uniformly filling the space between the layers. 

The use of a magnetization language for the response prompts for a magnetic field, and Maxwell's equations allow us to write $\bm{\hat{z}} \times (\bm E^{(2)} - \bm E^{(1)}) = \ii \, \omega \, a \langle \bm B_{\parallel} \rangle$,
where $ \langle \bm B_{\parallel} \rangle$ is the average parallel component  of the magnetic field between layers. It should be kept in mind, however, that the use of a magnetic language  for the response is merely a matter of convention, and the entire analysis can be carried out in terms of sheet currents and electric fields instead, see Eqs. (\ref{deltadiag}) and  (\ref{response}).

To second order in $\tfrac{\omega \, a}{c}\sim \tfrac{a}{\lambda}$, we can further replace the fields with their values at the nominal center of the bilayer, $ \bm E $ and $ \bm B $,
 %
%
and introduce a surface polarization density through $\bm j_T =\bm j^{(1)}+\bm j^{(2)}= -\ii \omega \bm p_{\parallel} $. The bilayer is then replaced by a  single sheet placed at $z_0=0$ with the standard volume polarization and magnetization $\bm P = \bm p_{\parallel} \delta (z),  \;\;\;  \bm M = \bm m_{\parallel} \delta (z)$, where $\delta(z)=\Theta((a/2)^2-z^2)/a$. The constituent equations of Eqs. (\ref{deltadiag}) and (\ref{response}) then read
\begin{equation}\label{const}
\begin{split}
\bm p_{\parallel} &= -2 \frac{\sigma_0+\sigma_1}{\ii \omega} \; \bm E_{\parallel} - a \sigma_{xy}  \;  \bm B_{\parallel} \\
\bm m_{\parallel}   &= a \sigma_{xy}  \; \bm E_{\parallel}  +  \ii \omega \frac{a^2}{2} (\sigma_0-\sigma_1) \;  \bm B_{\parallel}
\;,\end{split}
\end{equation}
The form of Eqs. (\ref{const}) is  often 
taken as the starting point
in discussions of chiral molecules.\cite{Barron04}
Here we have deduced them from the basic layer response of Eq. \ref{response}. The presence of the cross-term $\sigma_{xy}  $ leads to the optical activity in twisted bilayer graphene, as explained in Refs. \cite{Kim16,Suarez17}.

For 2D materials at optical frequencies, the last term is already of order $(\tfrac{a}{\lambda})^2$ and usually dropped.\cite{Expansion}  Here, we will keep it because at the neutrality point, the in-plane magnetic susceptibility may be the only response and it turns out to be huge for twisted bilayer graphene at the magic angle, for which the lowest band becomes almost flat.\cite{Bistritzer11} 

The basic content of Eq. (\ref{const}), that
 an in-plane magnetic field  leads to an electric current and a longitudinal current is accompanied by a magnetic moment, 
 remains valid at zero frequency for twisted bilayer. Upon doping this is also true for plasmons, which is one of the major results of this work: the presence of longitudinal plasmons that carry a magnetic moment.

{\it Static response.} 
In the context of optical activity, the response of twisted bilayer has been discussed for interband transitions at finite frequency.\cite{Kim16,Suarez17}
Here, we focus 
on the response at zero frequency including intraband transitions upon doping,
 i.e., the Drude component to the conductivity. 
The real part of the conductivity is given by $\Re\sigma_n(\omega)=\pi D_n\delta(\omega)+\Re\sigma_n^{reg}(\omega)$, with $n=\{0,1,xy\}$ and where the regular part of the conductivity $\sigma_n^{reg}$ is obtained from the usual Kubo formula applied to the twisted bilayer Hamiltonian using the continuous model of Refs. \cite{Lopes07,Bistritzer11}, see SI.\cite{SI} The imaginary part follows from the Kramers-Kronig relation. 

\begin{figure}
\includegraphics[width=0.99\columnwidth]{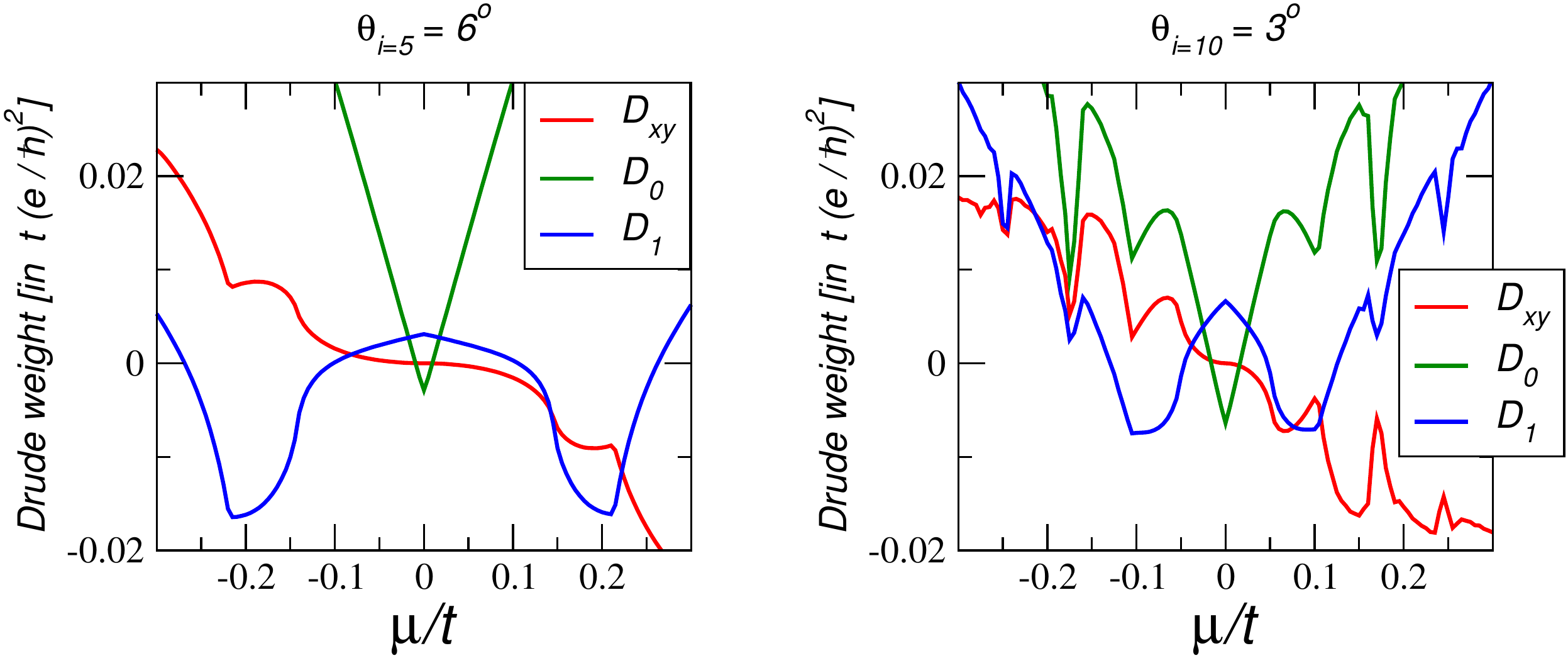}
\caption{(color online): The transverse Drude weight $D_{xy}$ in units of $t/\hbar^2$ as function of the chemical potential for two twist angles $\theta_{i=5} = 6^\circ$ (left) and $\theta_{i=10} = 3^\circ$ (right). Also shown the Drude weights $D_0$ (green) and $D_1$ (blue).  
\label{Dxy}}
\end{figure}

The Drude weight, which measures the density of inertia, is here a $4\times4$ Drude matrix with three components $D_{0,1,xy}$ defined by $D_n=-i\lim_{\omega\to0}\omega\sigma_n$. They are shown in Fig. \ref{Dxy} for two different twist angles labeled by $i$ with $\cos\theta_i=1-\frac{1}{2A_i}$ and $A_i=3i^2+3i+1$. All curves show discontinuities which are related to the band structure as already discussed in Ref. \cite{Stauber13} in the case of $D_{xx}=D_0+D_1$. In the following, we will discuss the results in detail.  

{\it Response at the neutrality point.}
At neutrality, there are no charge carrier and no total current can flow. Therefore, $D_0 +D_1 =0$ and also $D_{xy}=0$ at $\mu=0$. However, the counterflow\cite{Bistritzer11} $D_0(\mu=0)-D_1(\mu=0)=2D_0(\mu=0)$ does not have to be zero (and can even be negative), provoking a magnetic response to an in-plane magnetic field:
\begin{align}\label{counterflow}
\bm m_{\parallel}  =-a^2\frac{D_0}{1+\ii(\omega\tau)^{-1}}\bm B_{\parallel}\;,
\end{align}
where we also included a finite relaxation time $\tau$. At first glance, talking about counterflow when no free carriers are present seems bizarre. Nevertheless, in the clean limit $\tau\to \infty$, a finite value of $D_0$ in Eq. (\ref{counterflow}) merely expresses the emergence of a magnetic moment upon the adiabatic application of a magnetic field: a perfectly sensible and time-reversal invariant result for which no carriers need to be present, as neutral graphene attests.

Since $D_0<0$, it amounts to a paramagnetic response. It is important to note that again this effect is only possible for finite $a$ and would vanish in the limit of $a\to0$.
At the magic angle $i=31$, we find $D_0\approx-8.5e^2t/\hbar^2$, which corresponds to a response  about 200 times larger than the (diamagnetic) one of core electrons in bilayer graphene,\cite{DiSalvo79} see also SI.\cite{SI} This is an unprecedented in-plane magnetic response which is only related to the counter-flow of the two adjacent graphene layers. It is also much larger than the lattice response of single layer graphene\cite{Gomez11} or MoS$_2$\cite{Gutierrez16} due to a {\em perpendicular} magnetic field. 

{\it Hall response at finite doping.} At finite chemical potential, the longitudinal Drude weight $D_{xx}=2(D_0+D_1)$ is always positive
 and proportional to $\mu$ in the limit of $\mu\to0$, in agreement with the standard mass tensor result for conical bands. The Drude weight of the counterflow $D_{CF}=(D_0-D_1)$ should become positive for some finite $\mu$ and then yield a diamagnetic response as expected from Lenz's rule. Its proportionality with respect to the chemical potential was first given in Ref. \cite{Bistritzer11}, but the negative offset leading to the paramagnetic response at half-filling was not discussed.

$D_{xy}$ becomes finite for $\mu\neq0$ and shows ambipolar behavior similar to a Hall response, i.e., positive for say $\mu<0$ and negative for $\mu>0$, also reversing sign upon twist angle sign reversal. From Eq. \ref{const}, a finite $D_{xy}$ renders the longitudinal current to be accompanied by a magnetic moment which also holds for plasmonic excitations, as later explained. Furthermore, there is the possibility of a transverse response at $\omega=0$ and an in-plane magnetic field induces a longitudinal current at finite chemical potential 
in the clean limit. As in Eq. \ref{counterflow} , observation of this  {\em longitudinal} Hall response in the presence of dissipation would require a finite frequency with $\omega \tau\gg1$,
\begin{align}\label{longHall}
\bm  j_T  =-a\frac{D_{xy}}{1+\ii(\omega\tau)^{-1}}\bm B_{\parallel}
.\end{align}
 
{\it Intrinsic excitation.} Intrinsic excitations or plasmons are collective longitudinal and/or transverse current oscillations. In the limit $a\to0$, they were discussed for TBG in Refs. \cite{Stauber13,Stauber16} and the response only depends on $\sigma_0+\sigma_1$. Here we will investigate the influence of the transverse response $\sigma_{xy}$ for finite $a$.

Plasmons can be obtained in an elementary fashion from the constitutive relation Eqs. (\ref {deltadiag}) and (\ref{response}). In the instantaneous approximation, valid to order $(v/c)^2$, the self-fields are purely longitudinal and, decomposing the Fourier components of the current into longitudinal and transverse parts, $\bm j^{(n)}=j_n{\bf \hat q}+\delta j_n{\bf \hat q_\perp}$ with ${\bf \hat q_\perp={\bf \hat z}\times{\bf \hat q}}$, non trivial solutions for the currents are given by the zeros of the determinant of a $2\times2$-matrix:
\begin{align}\label{2by2} 
\det\left[{\bf 1}_{2\times2}- 
\begin{bmatrix}  \chi_0   & \chi_1   \\ \chi_1 & \chi_0  \end{bmatrix}\mathcal{D} \right] = 0
,\end{align}
where  $\chi_n = -\ii \omega \sigma_n$, with the photonic propagator of a double layer structure $\mathcal{D}$ defined in Ref. \cite{Stauber12} (see also SI).

Two branches appear in the limit  $q\to0$: the ordinary 2D plasmon with $\sqrt{q}$ dispersion and an acoustic one. Whereas the latter is vulnerable to the local approximation\cite{Stauber2012}, the ordinary plasmon, experimentally observed in single layer graphene\cite{Chen12,Fei12,Yan13}, is expected to be a robust feature. 
In the non-retarded approximation these longitudinal plasmon frequencies are not modified by the transverse coupling $\sigma_{xy}$, see Eq. (\ref{2by2}). 
Nevertheless, a finite value of $\sigma_{xy}$ adds a transverse component to the current, given in the limit $q\to0$ by the following relation between electric and magnetic dipoles oscillations:
\begin{equation}\label{constraint}
 {\bf \hat q}\cdot \bm m = \chi \; {\bf \hat q}\cdot \bm j_T
,\end{equation}
with  $\chi=\tfrac{a \sigma_{xy}}{2 (\sigma_0 + \sigma_1)}$. The ordinary plasmon carries total charge $\bm q \cdot \bm j_T \neq 0 $ and, by Eq. \ref{constraint}, also carries a longitudinal magnetic moment, the signature of chiral excitations.\cite{Rosenfeld1926,Barron04,Govorov10} In the relevant limit $\omega\to 0$, the plasmon magnetic content just becomes a real number involving Drude terms: $\chi_0=\tfrac{a D_{xy}}{2 (D_0 + D_1)} $ (see also SI). Therefore, the non-zero value of the chiral Drude term $D_{xy} $ at finite doping  gives plasmons a chiral character. This analysis can be made more general by studying the mixed spectral density of total charge $ {\bf \hat q}\cdot \bm j_T(\bm q) $ and parallel magnetic moment ${\bf \hat q}\cdot \bm m(\bm q)$, see SI.\cite{SI} 
\begin{figure}
\includegraphics[width=0.99\columnwidth]{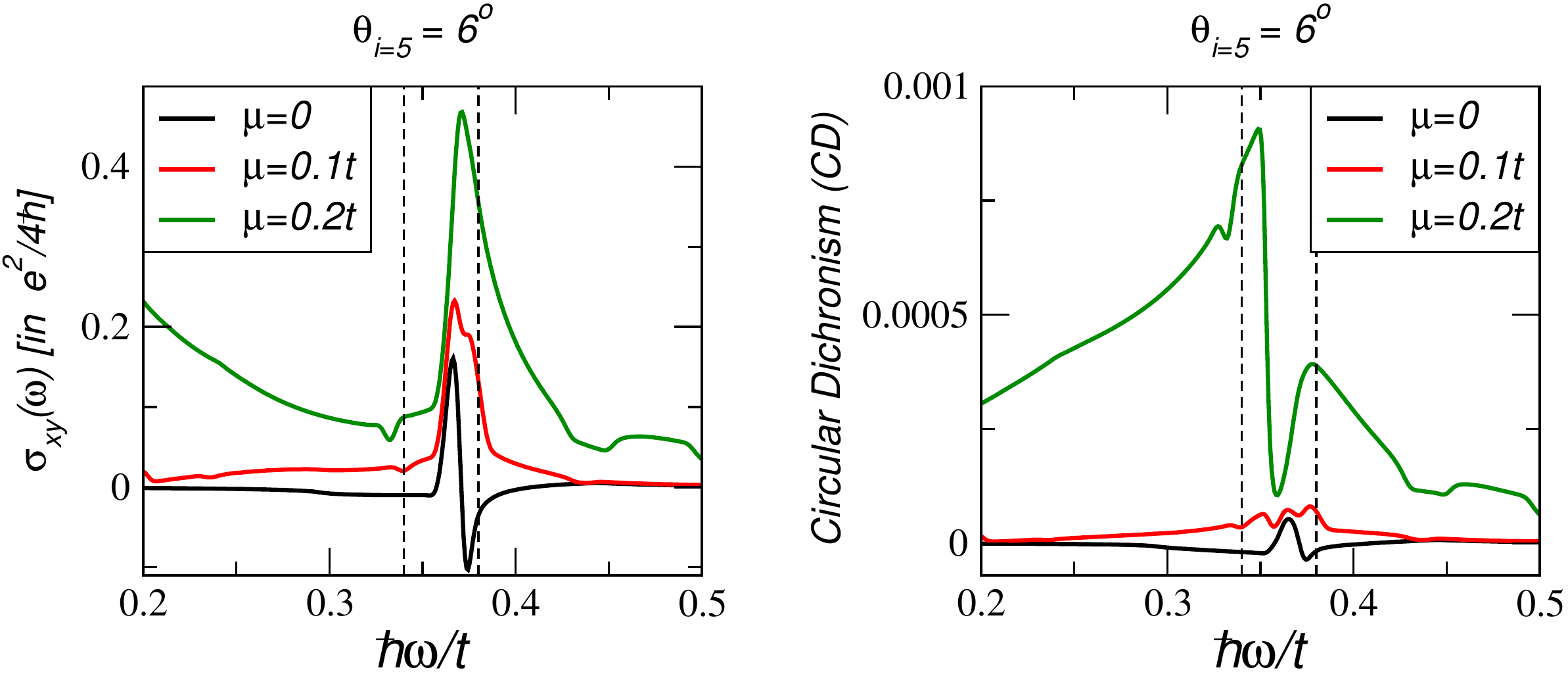}
\includegraphics[width=0.99\columnwidth]{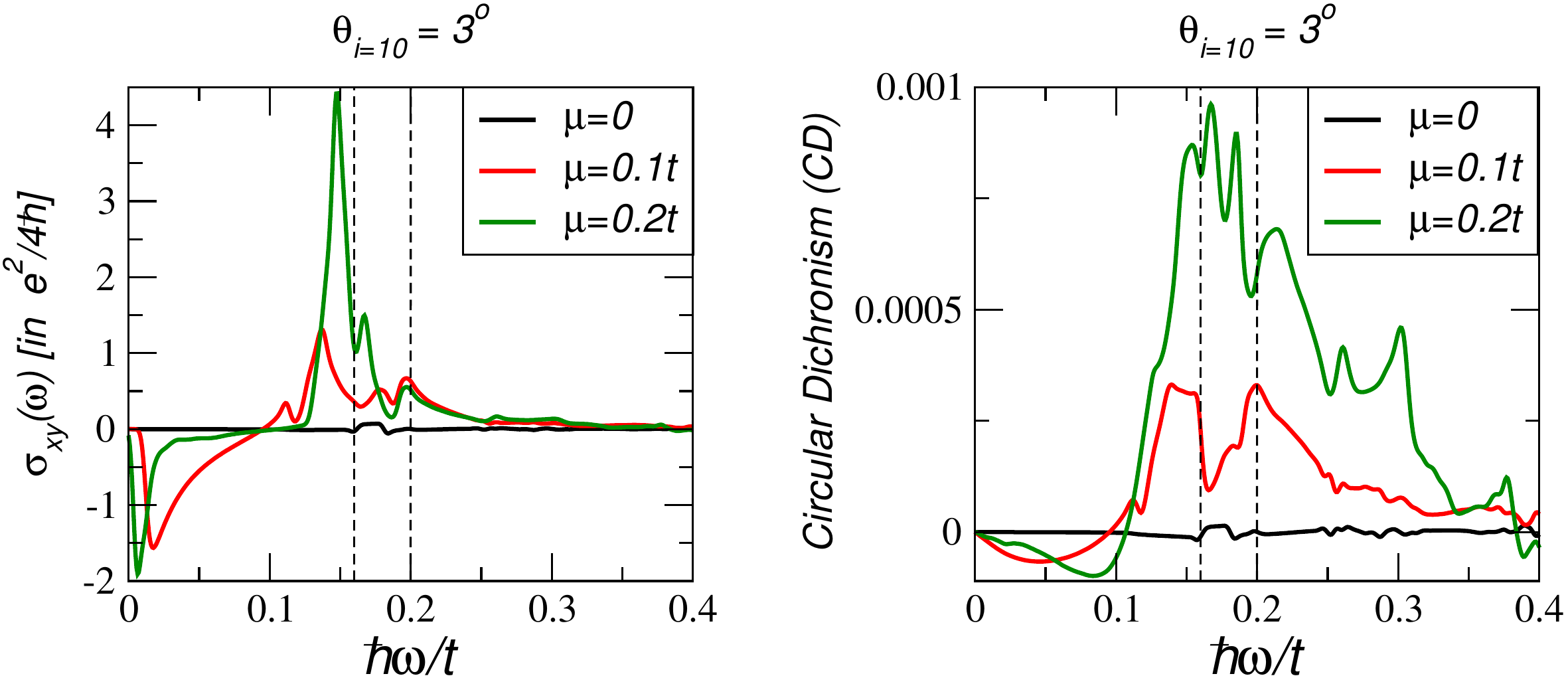}
\caption{(color online):  Left: The chiral or Hall conductivity at finite frequencies for several doping levels $\mu/t=0,0.1,0.2$ and the two angle $\theta=6^\circ$ (top) and $\theta=3^\circ$ (bottom). Right: The circular dichronism for several doping levels with $v_F/c=300$ and $\epsilon=1$ for the two angle $\theta=6^\circ$ (top) and $\theta=3^\circ$ (bottom). The vertical lines correspond the two energy scales $\epsilon_M=(2\pi/3)\hbar v_F/L_M$ and $\epsilon_{vH}\approx\epsilon_M-t_\perp/3$ with $L_M$ the Moir\'e-lattice period and $t_\perp$ the interlayer hopping amplitude, see also SI.
\label{CD}}
\end{figure}

{\it Optical activity.} The optical activity at finite frequencies can be significantly modified upon doping.  The circular dichroism (CD), defined as the relative difference in absorption for right and left-handed circularly polarized light,
$CD = \frac{\mathcal{A}_{+} - \mathcal{A}_{-}}{2 (\mathcal{A}_{+} + \mathcal{A}_{-})} $, can be written using the formalism of Eqs. (\ref{deltadiag}) and (\ref{response}) as follows, see SI:\cite{SI}  
\begin{equation}\label{Dichroism}
  CD = 
\frac{\Re (\sigma_{xy})}{2 \Re(\sigma_0+\sigma_1)} \sqrt{\epsilon}\;  \frac{\omega a  }{c} 
.\end{equation}
Apart from the presence of the dielectric, this formula essentially coincides with that of Ref. \cite{Kim16}. Notice that CD basically coincides with the factor $\chi$ of Eq. (\ref{constraint}), emphasizing its meaning as the chiral content of the transition. The key response $\Re( \sigma_{xy}) $ as well as the CD are plotted in  Fig. \ref{CD}  for several doping levels in the vicinity of the saddle frequencies, where optical activity is experimentally observed\cite{Kim16}. For zero doping, the results were first obtained for the present model in Ref. \cite{Suarez17} and compare well to the values observed in experiment. Its peak-dip  structure was linked to nearly cancelling opposite contributions and only the different chiral structure of the electrons in the two layers yields a finite Hall response. 

However, the chirality of the electrons is not the only source to break the cancelling symmetry in order to reach a finite Hall response. Also finite doping or bias\cite{SI} between the two layers yield a response - even assuming the same chirality of the electrons. In  Fig. \ref{CD}, we show how the $\mu=0$ bimodal structure evolves into a well-defined single peak structure upon doping, in addition to an overall increase in spectral weight. Incidentally, at small angles, the unimodal form of the doped case seems to better describe the experimental shape rather than the nominal zero-doping case.\cite{Kim16} 

\begin{figure}
\includegraphics[width=0.49\columnwidth]{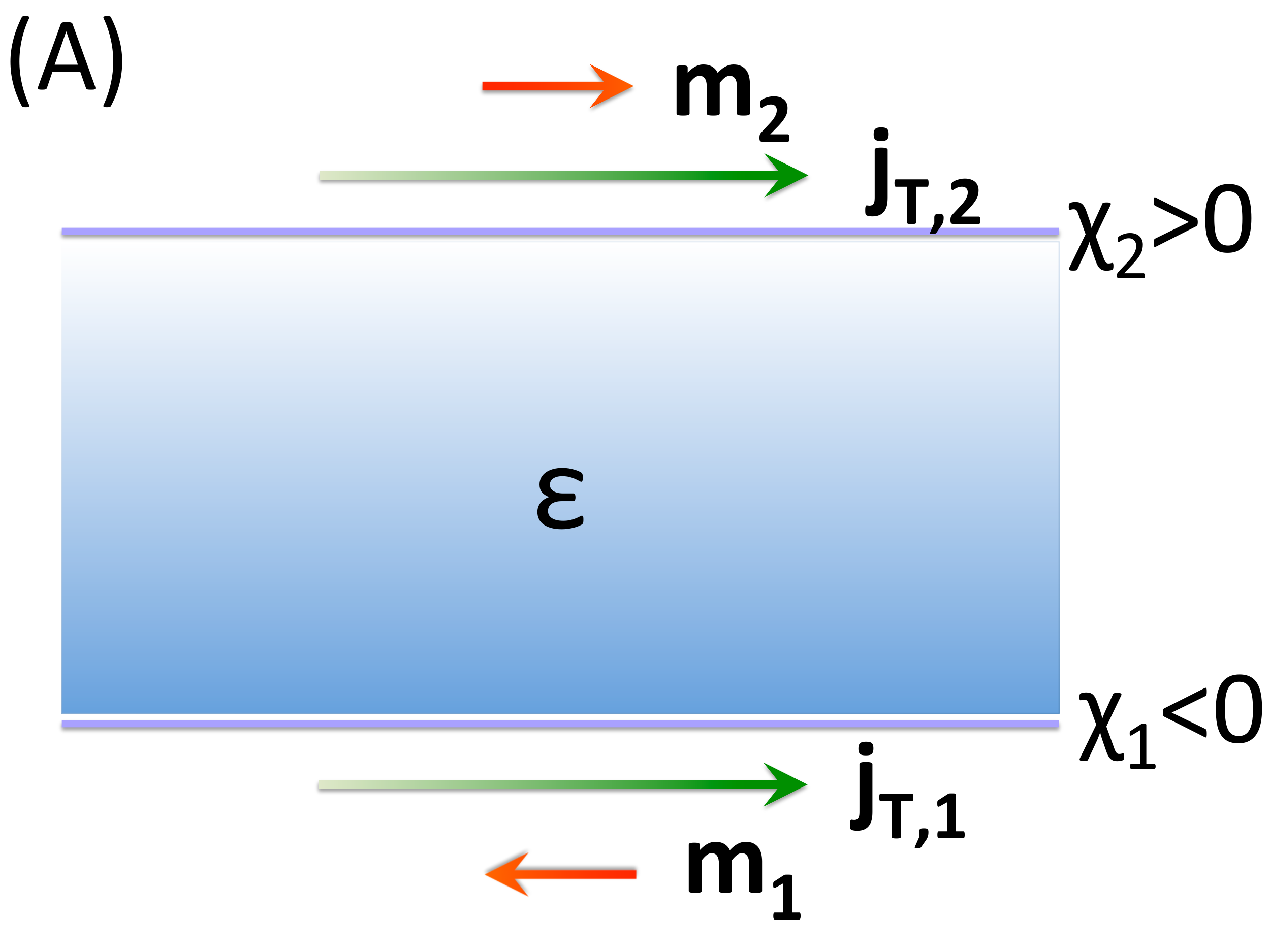}
\includegraphics[width=0.49\columnwidth]{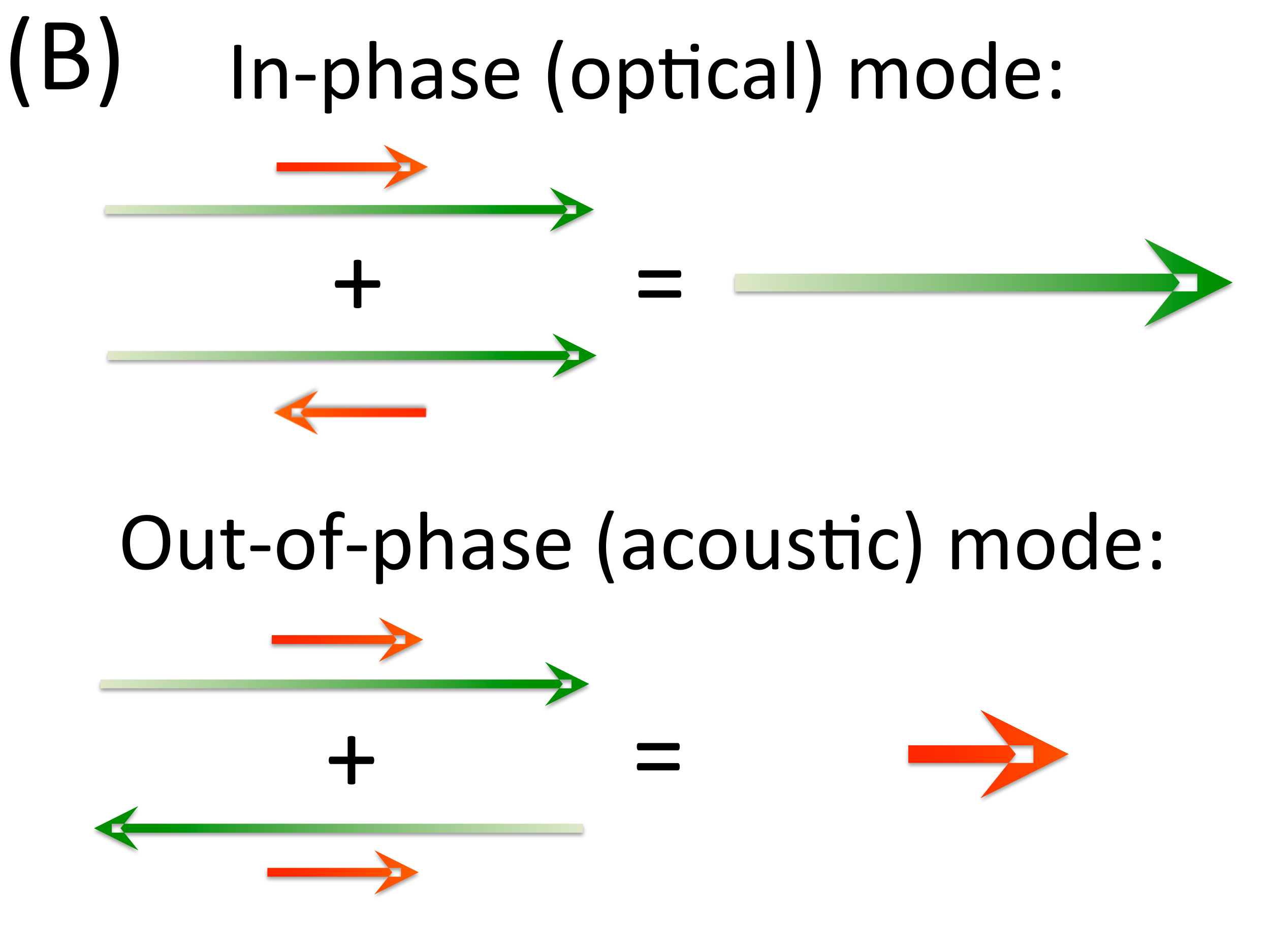}
\caption{(color online): (A) Double-layer structure with two "identical" twisted bilayer graphene samples, but  with opposite chiralities ($\chi_1=-\chi_2$) separated by an insulator with dielectric constant $\epsilon$. The current and attached magnetic moment are either parallel or antiparallel. (B) The two plasmon modes hybridize due to electrostatic interaction and form symmetric in-phase (optical) modes and anti-symmetric out-of-phase (acoustic) modes. The in-phase mode has purely charge-like character whereas the out-of-phase mode has a purely magnetic-like character.
\label{Magnetoplasmon}}
\end{figure}

{\it Discussion.} The electro-magnetic coupling in chiral materials opens up new scenarios for Hall physics if we can apply different (constant) gauge fields to the individual layers. Notice that these fields cannot be gauged away simultaneously. In Dirac systems such as twisted bilayer graphene this should be achieved by uniaxial strain which acts differently on the two layer e.g. by using a bended substrate. But strain also acts oppositely on the two K-points and it is thus necessary to break the symmetry between the $K$ and the $K'$-point which are related by time-reversal and parity symmetry. The breaking of the valley-symmetry can be achieved by valley-polarising the sample by an electrical current through a narrow constriction,\cite{Rycerz07} and we expect a transverse current if the sample is strained differently for the two layers. Valley polarisation can also be achieved by pumping, see Ref. \cite{Jiang13}.

One can further design purely magnetic plasmons by placing two twisted bilayer graphene structures with opposite chiralities at the two opposite faces of a dielectric substrate. Electrostatic interaction couples the plasmonic excitations on different faces and the out-of-phase oscillations do not carry charge, but the magnetic moments add up similar to what happens in 3D topological insulators, see Fig. \ref{Magnetoplasmon}.\cite{StauberTI13} The two modes can be addressed separately by changing the distance of the exciting electric dipole with respect to one of the surfaces.\cite{Ameen17}
 
{\it Conclusions.} By studying the most general local response function for a two-layered 2D system, we discussed general properties of optically active 2D systems. We applied our analysis to twisted bilayer graphene and calculated the intraband response (Drude weight) from the Kubo formula. In general terms, we predict novel phenomena related to the covalent drag $D_1$ and Hall-like $D_{xy}$ terms of the Drude matrix, i.e., a huge paramagnetic susceptibility at zero chemical potential due to counterflow, a longitudinal Hall effect that can be induced by uniaxial strain, and chiral magneto-plasmonic excitations, endowing plasmons and the associated near field fluctuations\cite{Tang10} with a chiral character. Moreover, the optical activity in terms of the circular dichronism can be considerably enhanced with doping.

Twisted plasmons accompanied by a magnetic moment were predicted to also exist in 3D topological insulators,\cite{Raghu10} but have been evasive in experiments so far. One of the reasons might be that the thin samples used in experiments (for thick samples the noise from the bulk increases) only carry optically active oscillations that are purely charge-like.\cite{StauberTI13} Chiral 2D materials thus offer a new platform to observe these novel chiral plasmons that can further be combined with other near-field chiralities due to spin-momentum locking.\cite{Canaguier15,Lodahl17}

Our response theory can be applied to any two-layered system with a rotational symmetry $C_{n>2}$. Optical activity then hinges on the existence of a finite interlayer $\sigma_{xy}$ (Hall) response, a feature expected for any layered system that does not coincide with its parity-reversed image, without the need for time-reversal breaking. Thus, twisted bilayer graphene might well be just one example among a potentially large class of layered materials where our work becomes relevant, i.e., we expect it to be present in any 2D van der Waals heterostructures with geometrical chiral structure. This opens up a new way to design novel chiral metamaterials without breaking time-reversal or rotational symmteries.

{\it Acknowledgements}. We acknowledge interesting discussions with Luis Brey. This work has been supported by Spain's MINECO under Grant No. FIS2017-82260-P, FIS2015-64886-C5-5-P, and FIS2014-57432-P and by the Comunidad de Madrid under Grant No. S2013/MIT-3007 MAD2D-CM. TL acknowledges support by the National Science Foundation NSF/EFRI grant (\#EFRI-1741660).

\newpage
\begin{widetext}
{\bf\huge Supplemental Information}
\section{Hamiltonian}

The non-interacting Hamiltonian used for calculating the response to total fields is given by\cite{Lopes07,Bistritzer11}
\begin{align}\label{Hamiltonian}
\mathcal{H } =&\hbar v_F\sum_{\bm k} c_{1,\bm k,\alpha}^{\dagger} \;\bm \tau_{\alpha\beta}^{-\theta/2} \cdot (\bm k + \frac{\Delta\bm K}{2}+\Delta) \; c_{1,\bm k,\beta} \notag\\ 
+&\hbar v_F\sum_{\bm k} c_{2,\bm k,\alpha}^{\dagger} \;\bm \tau_{\alpha\beta}^{+\theta/2} \cdot (\bm k - \frac{\Delta\bm K}{2}-\Delta) \; c_{2,\bm k,\beta}\\
+&t_{\perp} \sum_{\bm k,\bm G} (c_{1,\bm k + \bm G,\alpha}^{\dagger} \; T_{\alpha\beta}(\bm G) \; c_{2,\bm k,\beta} + H. c.)\notag\;,
\end{align}
where $( \bm \tau^{\gamma}_x,\bm \tau^{\gamma}_y ) =  e^{\ii\gamma\bm\tau_z/2} ( \bm \tau_x, \bm \tau_y ) e^{-\ii\gamma\bm\tau_z/2}  $, $\bm \tau_{x,y,z}$ being Pauli matrices. The separation between twisted cones is $\Delta \bm K = 2 |\bm K| \sin(\theta/2)  \left[ 0,1\right]$  with $\bm K = \tfrac{4 \pi}{3 a_g} \left[ 1,0\right] $. $\Delta$ introduces a potential difference between the two layers and is usually set to zero. Interlayer hopping is restricted to wavevectors $\bm G = \{\bm 0,-\bm G_1,-\bm G_1-\bm G_2\} $ with $\bm G_1 =  |\Delta \bm K| \left[ \tfrac{\sqrt{3}}{2},\tfrac{3}{2}\right]$, $\bm G_2 =  |\Delta \bm K| \left[ -\sqrt{3},0\right]$, and
\begin{equation}\label{Hopping}
T(\bm 0) = \begin{bmatrix} 1 & 1\\1 & 1\end{bmatrix}; \; \; T(-\bm G_1) = T^{*}(-\bm G_1-\bm G_2)=\begin{bmatrix} 
e^{\ii 2 \pi / 3} & 1\\ e^{-\ii 2 \pi / 3}& e^{\ii 2 \pi / 3}\end{bmatrix}
.\end{equation}
Calculations are performed with $t=2.78\,\text{eV}$ and $t_{\perp}=0.33\,\text{eV}$, being $\hbar v_F= \tfrac{\sqrt{3}}{2} t a_g $ with
 graphene lattice constant
$a_g=2.46 \, \mathring{\text{A}}$. The interlayer distance has been taken as $a=3.5 \, \mathring{\text{A}}$. 
 Also, twist angles have been chosen from the set of commesurate structures labelled by $\cos(\theta_i)=1-\tfrac{1}{2(3i^2 + 3i + 1)}$.

\section{ Linear response}
\subsection{Kubo formula}

The $4\times4$ conductivity tensor is 
\begin{equation}
\sigma^{(i,j)}_{\alpha\beta} = \ii \frac{\\e^2}{\omega_+} \, \chi_{j^{(i)}_\alpha\, j^{(j)}_\beta}(\bm q=0,\omega)
,\end{equation}
where $ \omega_+=\omega + \ii 0^+$, with axis indices $\alpha(\beta)=x,y $ and plane indices $i(j)=1,2$.

The retarded current-current response is given by 
\begin{equation}\label{Kubo}
 \chi_{j^{(i)}_\alpha\, j^{(j)}_\beta}(\bm q=0,\omega) = g_s g_v\int \frac{d^2 \bm k}{(2 \pi)^2} \sum_{n,m}
\frac{n_F(\epsilon_{m,\bm k}) - n_F(\epsilon_{n,\bm k})}{\omega_+ - \epsilon_{n,\bm k} + \epsilon_{m,\bm k}} 
\langle m,\bm k|j^{(i)}_\alpha |n,\bm k\rangle \langle n,\bm k|j^{(j)}_\beta |m,\bm k\rangle
.\end{equation}
Here, $g_s = g_v= 2$ are the spin and valley degeneracies. The states $|m, \bm k\rangle$ are eigenstates of $\mathcal{H}$
in subband $m$ and of momentum $\bm k$ in the first Brillouin zone of the superstructure. Their
eigenenergies are $\epsilon_{n,\bm k} $ and $n_F$ is the Fermi function. For graphene the current operator is 
$\bm j = e  v_F\bm \tau $.

\subsection{Symmetry considerations}

Here we review the symmetry arguments that lead to the  particular form of the response matrix and its dependence on twist angle. We base our discussion on $\bm \chi$ and, although calculations are based on the approximate model of Eq. \ref{Hamiltonian}, the conclusions apply generally for they are based on symmetry considerations. 

First of all, time reversal invariance makes $\bm \chi $  a symmetric matrix. Secondly, although real  graphene only exhibits three-fold rotational symmetry  around $\bf \hat z $, the perpendicular axis, this is enough to enforce full rotational symmetry for the $\bm q=0$ response considered here, a fact that remains true beyond the simplified continuum model. Full rotational symmetry requires 
 $\bm \chi$ to commute with the Pauli matrix 
$\bm \tau_y $, the generator of rotations in axis indices. Therefore, each of the four $2\times2$ submatrices making the full $4\times4$ matrix $\bm \chi$, should commute with $\bm \tau_y  $. This fact   together with its symmetric nature leads to 
\begin{equation}\label{SIresponse}
\bm \chi = 
\begin{bmatrix}  \chi_0^1 \bm 1 & \chi_1 \bm 1  -\ii\chi_{xy} \bm \tau_y \\ \chi_1 \bm 1+\ii\chi_{xy} \bm \tau_y & \chi_0^2 \bm 1 \end{bmatrix}
,\end{equation}
the form presented in the main text for the conductivity matrix, with the translation $\bm \sigma = \tfrac{\ii}{\omega} \bm \chi$.

Concerning parity relations for the twist angle $\theta$, it suffices to realize that a $\pi$ rotation of each graphene layer around an in-plane axis, say $\bf \hat y$, while keeping each plane in its original position (notice that this is   {\it not a global} $\pi$ rotation that would also exchange layer positions) implies the following changes 
\begin{align}
\theta &\rightarrow - \theta \\
j_{x}^{(1,2)} &\rightarrow  -j_{x}^{(1,2)}\\
j_{y}^{(1,2)} &\rightarrow  +j_{y}^{(1,2)}
.\end{align}
Therefore, from  Eq. \ref{Kubo} for instance, this leads to 
\begin{align}
\chi_0^{(1,2)}(\theta) &= \chi_0^{(1,2)}(-\theta)\\
\chi_1(\theta) &= \chi_1(-\theta)\\
\chi_{xy}(\theta)& = -\chi_{xy}(-\theta)
,\end{align}
the relations quoted in the main text for the conductivity entries, again with the translation $\sigma = \tfrac{\ii}{\omega}  \chi$ for each entry.

Let us finally comment on the question if there is some simple (semiclassical) formalism for the chiral response $\sigma_{xy}$ which might be useful for modelling TBG based optical active components. As shown above, the optical activity relies on the lack of spatial inversion leading to a finite $\sigma_{xy}$. At the model Hamiltonian level of Eq. (\ref{Hamiltonian}), it is the twist angle dependence on the interlayer hopping where this symmetry is broken. The relevant matrix elements in the linear response correspond to virtual processes where an 
electron-hole pair is created in one layer and destroyed in the other. 
Therefore, apart from rather obvious facts that the effect should be (at 
least) $\propto t_\perp^2$ and more prominent for frequencies around  the 
saddle points, where interlayer hopping is larger as experiments confirm \cite{Kim16}, little 
more can be said. The fact that seemingly innocuous approximations, e.g., neglecting the twist in some part of the TBG-Hamiltonian of Eq. (\ref{Hamiltonian}) as done in Ref. \cite{Bistritzer11}, can wash out the optical activity at zero doping \cite{Suarez17}, makes us believe that a simple recipe for characterizing $\sigma_{xy}$ by some simple formalism is not within reach. This would also fit the idea that $\sigma_{xy}$ is the outcome of nearly canceling contributions as discussed in Ref. \cite{Suarez17}.

\subsection{Results}
The current response to vector potentials is provided by $ \bm \chi $. Its 
calculation  involves the explicit evaluation of the imaginary part and Kramers-Kr\"onig recosntruction of the real part, with due care for the ultraviolet limit as explained in Ref. \cite{Stauber13}. The Drude matrix  corresponds to  $  \bm \chi(\omega\to0) $, a real quantity. It gives, in particular, the currents induced by the adiabatic application of a uniform, in-plane magnetic field, corresponding to constant though opposite vector potentials in each plane, with the results explained in the main text.

\begin{figure}
\includegraphics[width=0.99\columnwidth]{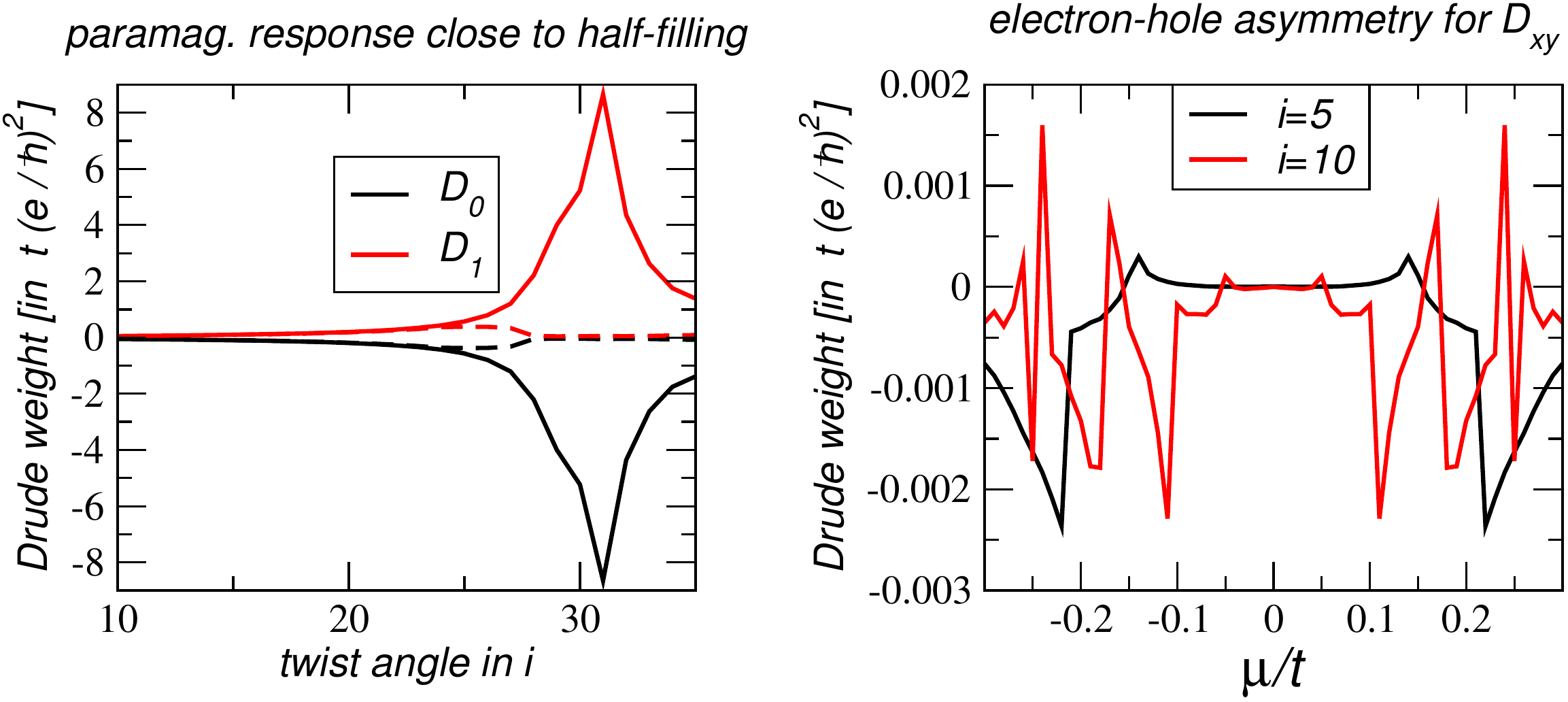}
\caption{(color online): Left hand side: The Drude weights $D_0$ (black) and $D_1$ (red) at half-filling (solid line) and $\mu=0.001t$ (dashed) as function of the twist angle $i$. Right hand side: The sum of the transverse Drude weight $D_{xy}(\mu)+D_{xy}(-\mu)$ as function of the chemical potential.
\label{FigAngle}}
\end{figure}

\begin{figure}
\includegraphics[width=0.99\columnwidth]{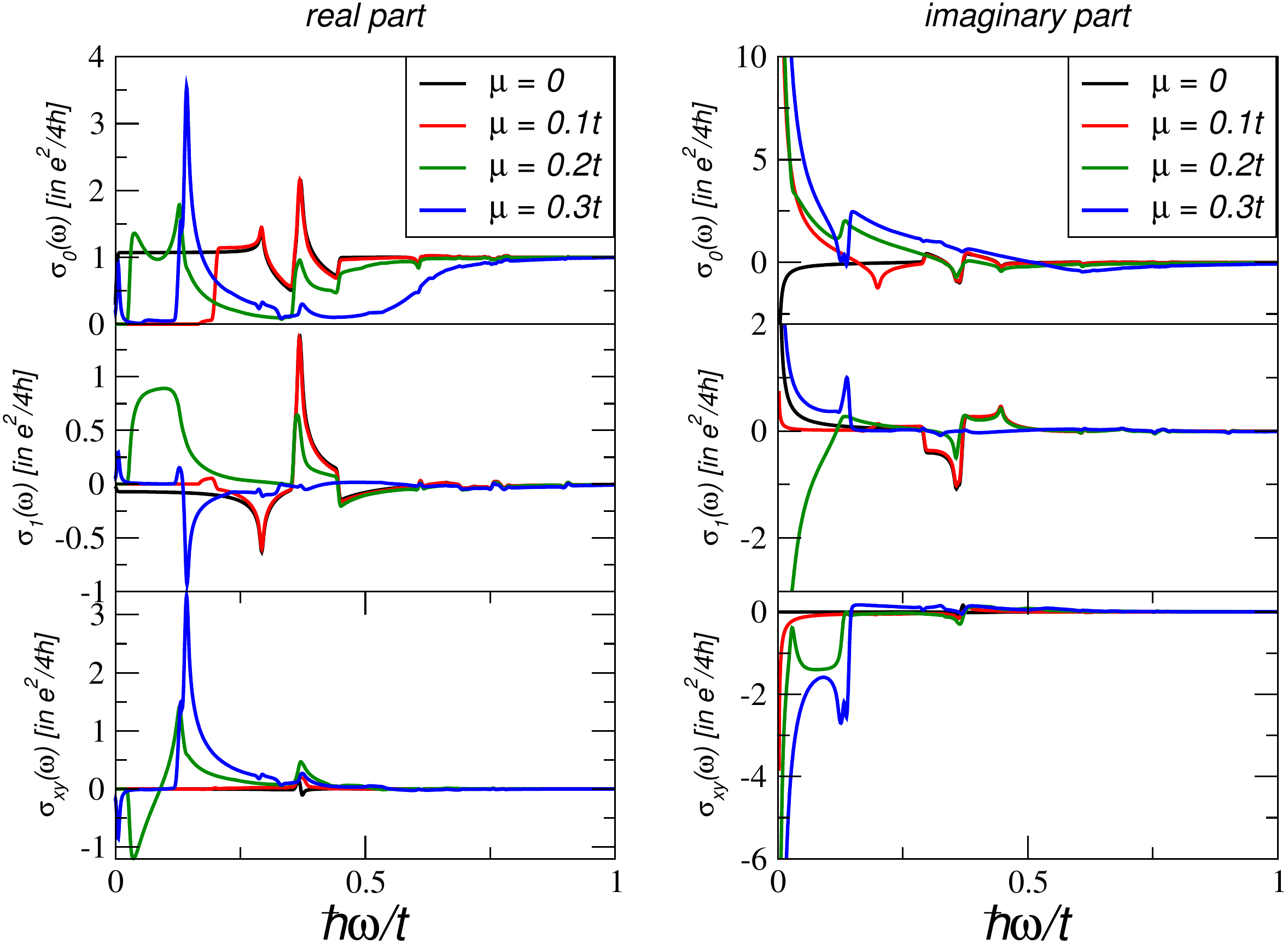}
\caption{(color online): The real (left) and imaginary (right) part of the several conductivities for a twist angle $\theta_{i=5} = 6^\circ$ at various chemical potentials.
\label{Chi}}
\end{figure}

Let us first dicuss the intraband (Drude) response in more detail. In Fig. \ref{FigAngle} (left), we show the Drude weight $D_0$ (black) and $D_1$ (red) at half-filling for different angles parametrised by $i$. $D_0+D_1=0$ within the numerical precision, but $D_0-D_1$ is always negative and minimal for the first magic angle with $i=31$. For a small electronic doping of $\mu=0.001t$, the strong paramagnetic response around $i=31$ is lost which can be used as a sensor (dashed curves). 

We can now compare the magnitude of the paramagnetic response of twisted bilayer graphene induced by the counterflow. The susceptibility at the neutrality point is given by  $\chi_{CF}/\mu_0=-a^2D_0$ with $\mu_0$ the magnetic permeability. Core electrons, not considered in our Hamiltonian, are another source of (dia)magnetic response. The estimate for single-layer graphene of Ref. \cite{DiSalvo79}, $\chi_{core}\sim4.8\times10^{-6}$emu/mol translates into $\chi_{core}/\mu_0=-0.02a^2e^2t/\hbar^2$ with $a=3.5 \,\mathring{\text{A}}$. For $i=31$, we have $D_0=-8.65e^2t/\hbar^2$; this means that for a bilayer, the counterflow response is about 200 times larger then the intrinsic diamagnetic response.

In Fig. \ref{FigAngle} (right), we display the electron-hole asymmetry by plotting the chiral Drude weight for both polarities, $D_{xy}(\mu)+D_{xy}(-\mu)$. Strain applied differently on the two layers acts as an effective magnetic field at one valley. Circularly polarized light only excites electron-hole pairs of one valley and an external source-drain biased will generate a current because the electronic current is not completely cancelled by the hole-like current. This can be the basis of a strain-induced photodetector.

Finally, we will discuss the full optical response at finite chemical potential. In Fig. \ref{Chi}, we present the three different conductivities at various chemical potentials for a twist angle with $i=5$. The chiral response $\sigma_{xy}$ becomes stronger for increasing chemical potential not only for transitions around the $M$-point at $\hbar\omega\approx0.35t$ as discussed in the main text,
but also at at lower frequencies $\hbar\omega\approx0.2t$, where new and very strong bands emerge.

\section{Chiral plasmons}

\begin{figure}
\includegraphics[width=0.59\columnwidth]{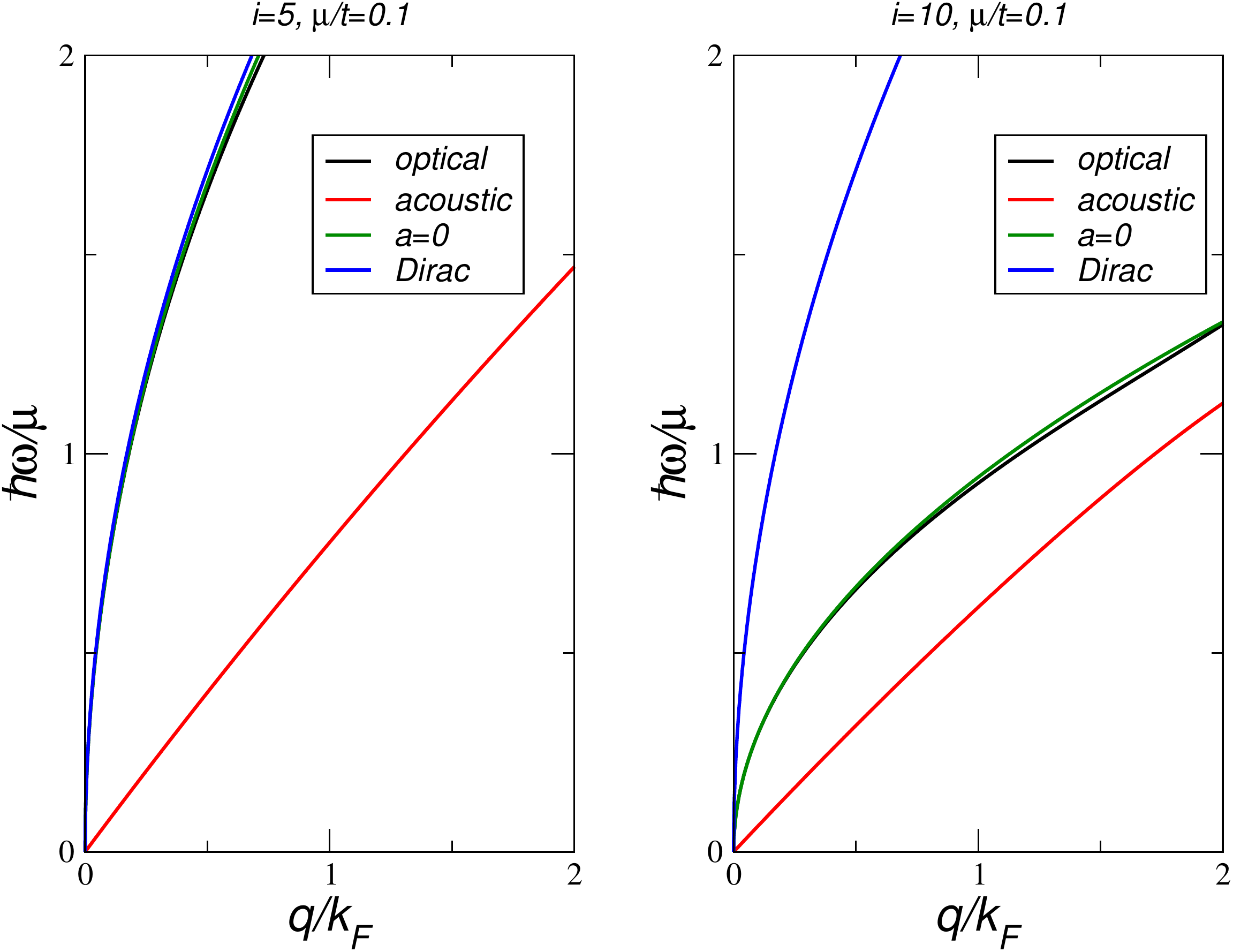}
\caption{(color online): Optical and acoustic plasmon dispersion for two twist angles $\theta_{i=5} = 6^\circ$ (left) and $\theta_{i=10} = 3^\circ$ (right). Also shown the plasmon dispersion for $a=0$ and for a pure Dirac system. 
\label{Plasmons}}
\end{figure}

 Coulomb interactions in the unretarded approximation are given by  the $4\times4$ photonic propagator, written in axis and plane indices as
$\mathcal{D}^{(i,j)}_{\alpha\beta} = \tfrac{q_\alpha q_\beta}{q^2}  \mathcal{D}(i,j)$, with
\begin{align}
\mathcal{D}(i,j) = 
\frac{q\epsilon_2}{\varepsilon_0\omega^2N}
\left(
\begin{array}{cc}
\cosh(qa)+\frac{\epsilon_3}{\epsilon_2}\sinh(qa)  & 1\\
1& \cosh(qa)+\frac{\epsilon_1}{\epsilon_2}\sinh(qa)
\end{array}\right) 
,\end{align}
where $N_=\epsilon_2(\epsilon_1+\epsilon_3)\cosh(qa)+(\epsilon_1\epsilon_3+\epsilon_2^2)\sinh(qa)$ and $\epsilon_n$ are the dielectric constants surrounding the double-layer structure. According to the main text, the plasmons appear as self-consistent solutions of  
\begin{align}\label{2by2} 
\det\left[{\bf 1}_{2\times2}- 
\begin{bmatrix}  \chi_0   & \chi_1   \\ \chi_1 & \chi_0  \end{bmatrix}\mathcal{D} \right] = 0
,\end{align}
where  $\chi_n = -\ii \omega \sigma_n$. They are given explicitly for $\epsilon_{1,2}=1$ and $\epsilon_{3}=2$ by 
\begin{align}
\omega^2=\frac{B}{2A}\pm\frac{\sqrt{B^2-4AC}}{2A}\;
,\end{align} 
 where $A=\epsilon+1$, $B=[(3+\epsilon)/2+(1-\epsilon)/2e^{-2qa}]d\chi_0+2e^{-qa}d\chi_1$ and $C=[1-e^{-2qa}]d^2(\chi_0^2-\chi_1^2)/2$, with  $d=d_0 t_0$, $t_0=2((1+\sqrt{\epsilon})$ and $d_0 = \tfrac{q}{2 \epsilon_0 \omega^2} $.

The simplest discussion of plasmons is based on replacing the full (complex) response by its real Drude weight. There is a strong dependence of the Drude weight on the chemical potential and twist angle and this dependence is inherited by the plasmonic resonance. The dependence on the twist angle is shown as example in Fig. \ref{Plasmons}. 

The fact that plasmon frequencies do not depend on the chiral term $\sigma_{xy}$ in the instantaneous approximation is no accident, and can be shown to extend to charge-charge excitations. Nevertheless, the chiral term adds a magnetic moment to the plasmon, as explained in the main text. This  analysis can be made more general and rigorous studying spectral densities. 
The simultaneous content of magnitudes $A$ and $B$ for excitations at frequency $\omega$ is given by the spectral density $\rho_{AB}(\omega)$, 
\begin{equation}\label{rhoAB}
 \rho_{AB}(\omega)= 
\sum_{n,m} (P_n-P_m) \langle n|A|m\rangle \langle m|B^{\dagger}|n\rangle \delta (\omega +\omega_{m} - \omega_{n})
,\end{equation}
where the sum runs over eigenstates of the total Hamiltonian (including interactions),  with energies $\hbar \omega_{n(m)}$ and probabilities $P_{n(m)} $. For the present case where the operators are sheet currents and within the RPA approximation, they can be obtained from the imaginary part of the appropriate entries of the  response to {\it external} fields, $\bm \chi_{ext}$,  related to the response to the {\it total} field , $\bm \chi=-\ii \omega \bm \sigma$, by the usual expression 
\begin{equation}\label{RPA}
 \bm \chi_{ext}= 
\left[{\bf 1}-\bm\chi \mathcal{D}\right]^{-1} \bm \chi
,\end{equation}
where $\mathcal{D}$ is here the complete ($4\times4$) photonic propagator.

\begin{figure}
\includegraphics[width=0.49\columnwidth]{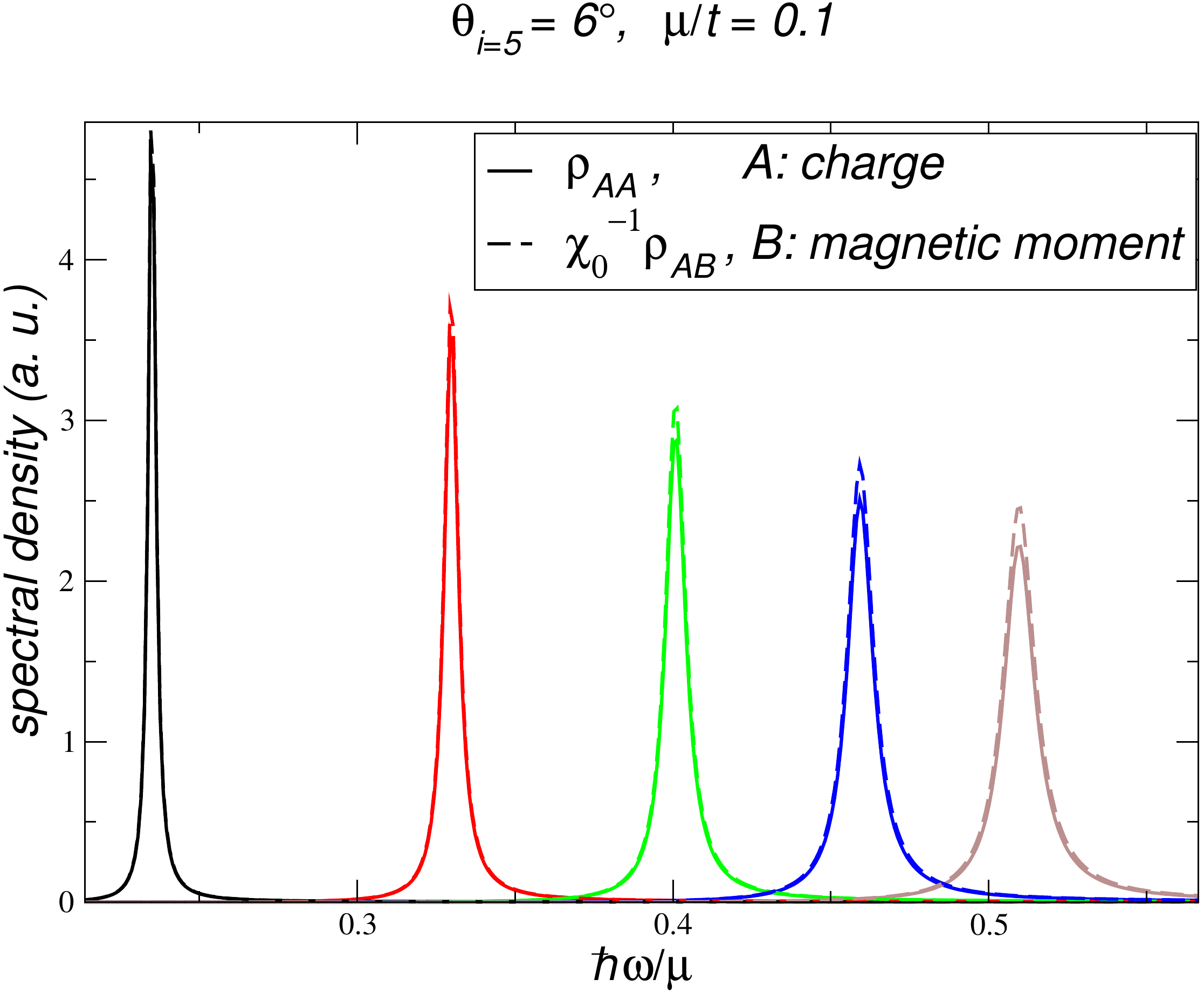}
\includegraphics[width=0.49\columnwidth]{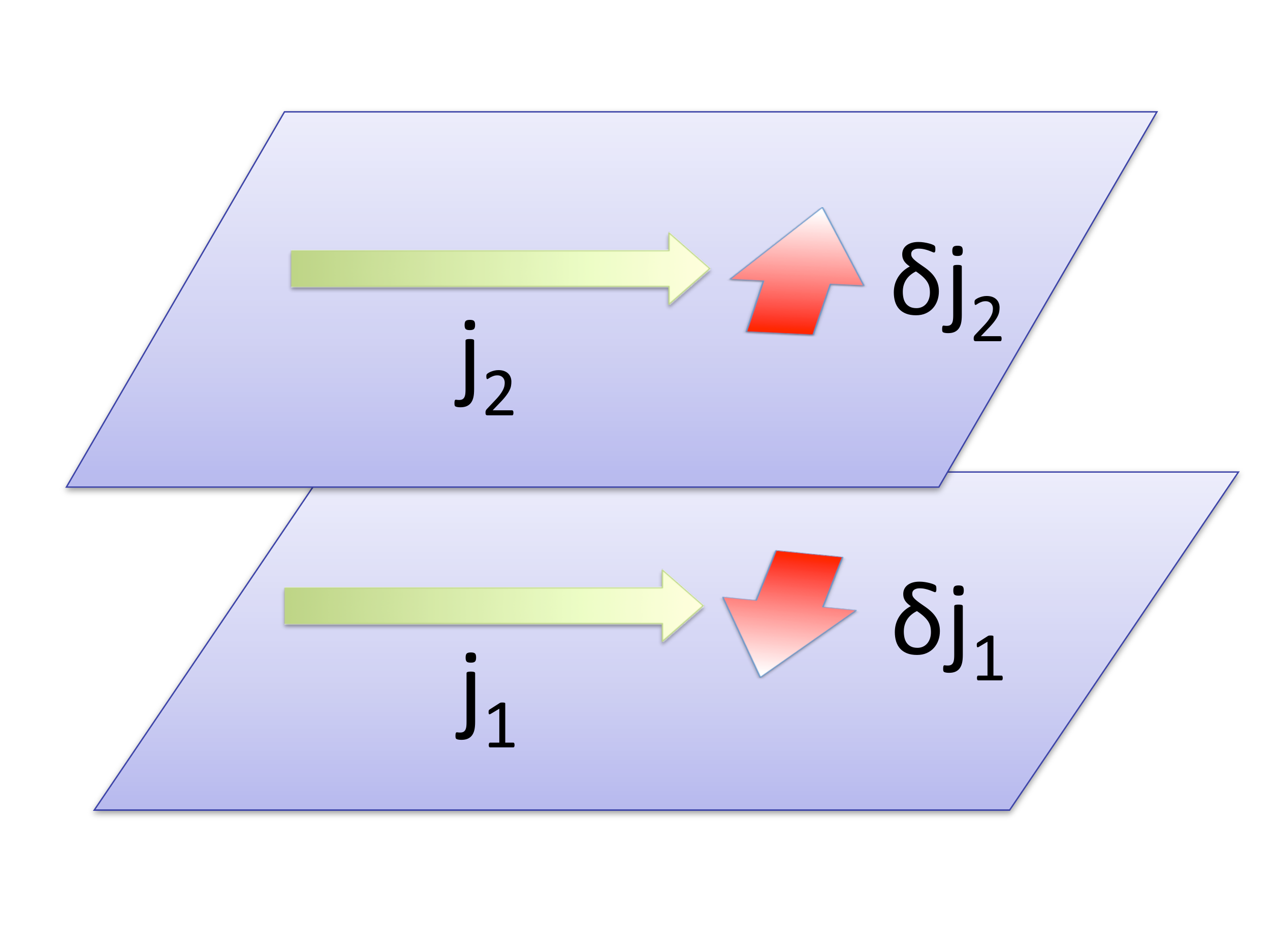}
\caption{(color online):  Left: Charge-charge spectral function (continuous lines)  and  charge-longitudinal magnetic moment scaled by $\chi_0$ (dashed lines) for five values of $q/k_F=
 0.01,0.02,0.03,0.04,0.05$ (left to right). Spectral poles have been made visible with a phenomenological damping $\tau \mu /\hbar=20$.
Right: Illustration of charged plasmon currents in twisted bilayer graphene decomposed into longitudinal and transverse components: $\bm j^{(n)}=j_n{\bf \hat q}+\delta j_n{\bf \hat q_\perp}$. A magnetic moment  is generated by transverse currents which are opposite in the two layers.}
\label{figspectral}
\end{figure}

The longitudinal nature of $\mathcal{D} $ makes the spectral densities mixing charge excitations, ${\bf \hat q}\cdot \bm j^{(1,2)} $,  not to depend on the chiral term of the response, in agreement with the elementary analysis before. In particular, plasmon poles at frequencies  given by the $2\times2 $ problem of Eq. \ref{2by2} appear in the corresponding spectral densities. On the other hand, a non-zero $\sigma_{xy} $ makes non-zero the mixed spectral density of total charge A: $ {\bf \hat q}\cdot \bm j_T(\bm q) $, and parallel
magnetic moment B:  $ {\bf \hat q}\cdot \bm m(\bm q)  $. This is shown in  Fig. \ref{figspectral}, where the charge-charge, {\it plasmon} spectral density $\rho_{AA} $, and mixed {\it plasmon-magnetic moment} spectral density $\rho_{AB} $, as obtained from the appropriate  entries of Eq. \ref{RPA}, are shown to nearly coincide when the latter is scaled by the factor  $\chi_0=\tfrac{a D_{xy}}{2 (D_0 + D_1)} $. Therefore, in agreement with the discussion in the main text, $\chi_0$ represents the magnetic content of the plasmon along the current direction \cite{chi}.

\section{Optical activity}

We assume the bilayer in vacuum but in nominal contact with a non-absorbing substrate occupying the half-space $z > 0⁺ $, and characterized by a light velocity $c_2=c/\sqrt{\epsilon}$. For an incident field given by $\bm A_{inc} = \bm{\hat{x}} \, A_o \, e^{\ii \frac{\omega}{c} z} $, the external fields acting on the bilayer  are
\begin{equation}
\bm E_{\parallel} = \ii \omega (1+r_0) A_0  \bm{\hat{x}}, \;\; \;\;\;\;\bm B_{\parallel} = \ii \frac{\omega}{c} (1-r_0) A_0 \bm{\hat{y}}
,\end{equation}
where $r_0$ and $t_0$ are the reflection and transmission amplitudes for the vacuum-substrate interface in the absence of graphene, given by 
\begin{equation}
r_0=\frac{c_2-c}{c_2+c}, \; \; t_0 = 1+r_0
.\end{equation}
To lowest order in the response, the fields produced by the induced polarization and magnetization (see constitutive equations in main text)  are now
\begin{multline}
\bm A_{\bm p,ind} = (- \frac{\sigma_0+\sigma_1}{ \epsilon_0 c} \, (1+r_0)^2  \;   \bm{\hat{x}} \; 
- \frac{\ii \omega a \sigma_{xy}}{2 \epsilon_0 c^2} \, (1-r_0^2)  \;   \bm{\hat{y}}) 
\times \begin{cases}
A_o \,e^{-\ii \frac{\omega}{c} z} &  z < 0,\\
A_o \,e^{+\ii \frac{\omega}{c_2} z} & z> 0,
\end{cases}
\end{multline}
and 
\begin{equation}
\bm A_{\bm m,ind} = - \frac{\ii \omega a \sigma_{xy}}{2 \epsilon_0 c^2} \, (1+r_0)  \;   \bm{\hat{y}} \; 
\begin{cases}
(-1+r_0) A_o \,e^{-\ii \frac{\omega}{c} z} &  z < 0,\\
(+1+r_0) A_o \,e^{+\ii \frac{\omega}{c_2} z} & z> 0.
\end{cases}
.\end{equation}

Using rotational invariance, the complete transmission and reflection amplitude matrices can be extracted from the previous results, leading to 
\begin{equation}
\begin{split}
t_{yx} &= -t_{xy} = - \frac{\ii \omega a \sigma_{xy}}{ \epsilon_0 c^2} \; (1+r_0)\\
t_{xx} &=  t_{yy} = (1+r_0) - \frac{\sigma_0+\sigma_1}{\epsilon_0 c} \; (1+r_0)^2
,\end{split}
\end{equation}
and 
\begin{equation}
\begin{split}
r_{yx} &= -r_{xy} = 0 \\
r_{xx} &=  r_{yy} = r_0 - \frac{\sigma_0+\sigma_1}{\epsilon_0 c} \; (1+r_0)^2 
.\end{split}
\end{equation}
 For the case of circularly polarized incident light, with polarization 
$\bm{\hat{u}}_{\pm} = (\bm{\hat{x}}  \pm  \ii \bm{\hat{y}}) / \sqrt 2$, the corresponding amplitudes are 
\begin{equation}
t_{\pm}  = (1+r_0) \; [1 - \frac{\sigma_0+\sigma_1}{\epsilon_0 c} \; (1+r_0) 
\mp  \frac{ \omega a \sigma_{xy}}{ \epsilon_0 c^2}] \, \bm{\hat{u}}_{\pm}
,\end{equation}
and 
\begin{equation}
r_{\pm}  =  [r_0 - \frac{\sigma_0+\sigma_1}{\epsilon_0 c} \; (1+r_0)^2 ] \, \bm{\hat{u}}_{\pm}
.\end{equation}

The  absorption is then  $\mathcal{A}_{\pm} = 1 - R_{\pm} - T_{\pm} $,  with the reflection and transmission coefficients (power)  given by $T_{\pm}=|t_\pm{}|^2 \tfrac{c}{c_2} $ and 
$R_{\pm}=|r_\pm{}|^2 $, leading  to the following expression for the  circular dichroism of twisted bilayer
\begin{equation}\label{Dichroism}
  CD = \frac{\mathcal{A}_{+} - \mathcal{A}_{-}}{2 (\mathcal{A}_{+} + \mathcal{A}_{-})} =
\frac{\Re \sigma_{xy}}{2 \Re(\sigma_0+\sigma_1)} \;  \frac{\omega a \sqrt{\epsilon}}{c} 
,\end{equation}
already presented in the main text.

An alternative characterization of optical activity is the rotation of the polarization plane of the transmitted field. For an incident plane wave linearly polarized along $\bm{\hat{x}}$, the transmitted field is proportional to 
$t_{xx} \bm{\hat{x}} + t_{yx} \bm{\hat{y}}$, and the polarization properties of the transmitted field are encoded in the ratio $R = \tfrac{t_{yx}}{t_{xx}} $, given to lowest order by 
\begin{equation}
 R = \frac{t_{yx}}{t_{xx}} \sim - \frac{\ii \omega a \sigma_{xy}}{ \epsilon_0 c^2}
.\end{equation}
Only for real $R$  is the transmitted light truly linearly polarized. For twisted bilayer graphene, $R$ is complex leading to an elliptically polarized transmitted field. Nevertheless, one can define the polarization rotation, $\Theta   $, as the angle formed by the long axis of the ellipse with respect to the $\bm{\hat{x}}  $ axis. In the present case, it is given by 
\begin{equation}
\Theta = \operatorname{Re}\,(\frac{t_{yx}}{t_{xx}}) =  \frac{ \omega a}{c} \;\frac{ \Im \sigma_{xy}}{ \epsilon_0 c}
,\end{equation}
to lowest order in the graphene response.

\section{Interlayer bias}
We apply a finite bias between the two layers labeled by $\Delta$ in Eq. (\ref{Hamiltonian}) which can model a perpendicular electric field.\cite{McCann06} This leads to different conductivities in the two layers and to additional terms in the constitutive equations. With $2\bar\sigma_0=\sigma_0^1+\sigma_0^2$ and $2\sigma_0^-=\sigma_0^1-\sigma_0^2$, they now read: 
\begin{align}\label{const}
\bm p_{\parallel} &= -2 \frac{\bar\sigma_0+\sigma_1}{\ii \omega} \; \bm E_{\parallel} - a \sigma_{xy}  \;  \bm B_{\parallel} -a\sigma_0^-\bm{\hat{z}}\times \bm B\\
\bm m_{\parallel}   &= a \sigma_{xy}  \; \bm E_{\parallel}  +  \ii \omega \frac{a^2}{2} (\bar\sigma_0-\sigma_1) \;  \bm B_{\parallel}-a\sigma_0^-\bm{\hat{z}}\times \bm E 
\;. 
\end{align}

A interlayer bias alters the response of the Hall conductivity and is finite even if the twist induced chirality of the electrons is not changed. In Fig. \ref{Delta}, we show the results for two different twist angles for twisted bilayer graphene at zero chemical potential. 

\begin{figure}
\includegraphics[width=0.99\columnwidth]{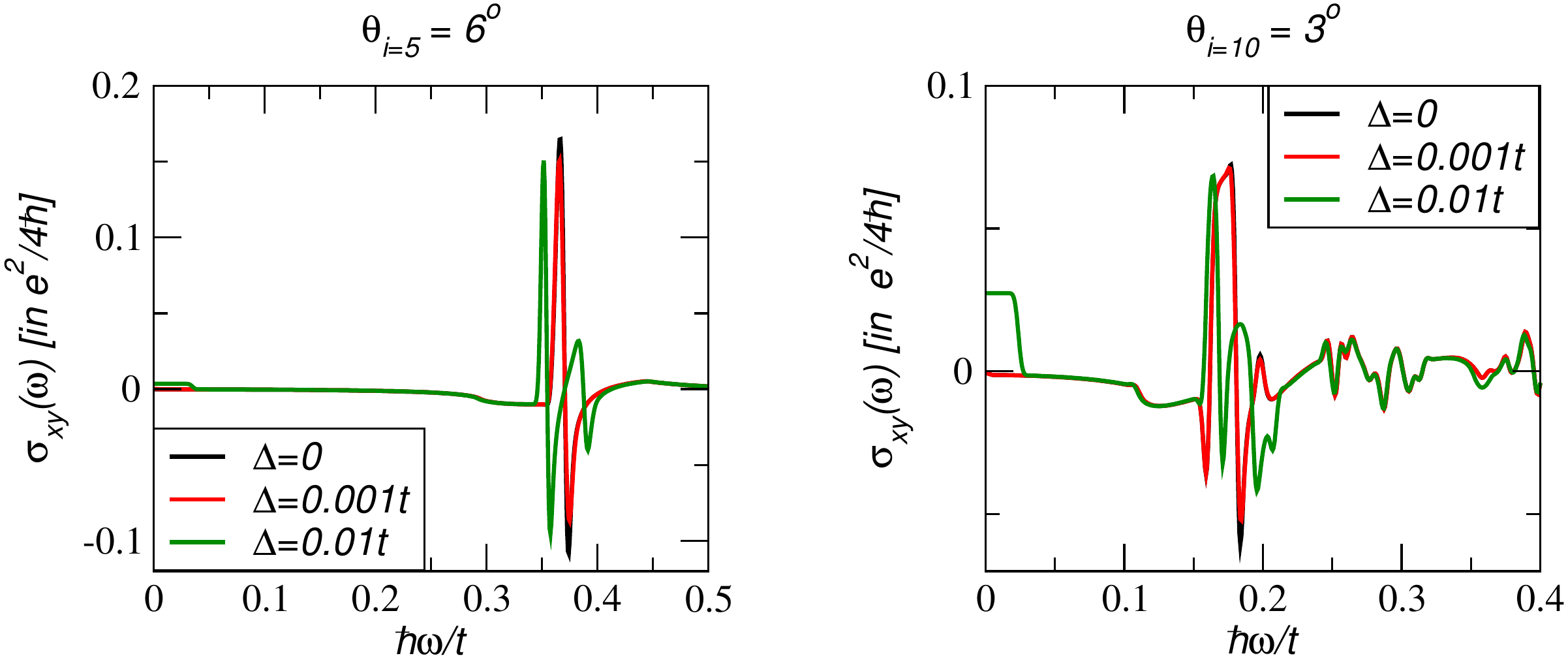}
\caption{(color online): The Hall conductivity $\sigma_{xy}$ at the neutrality point $\mu=0$ for different interlayer bias $\hbar v_F\Delta/t=0,0.001,0.01$ for the twist angles $\theta_{i=5} = 6^\circ$ and $\theta_{i=10} = 3^\circ$.
\label{Delta}}
\end{figure}

\end{widetext}
\bibliography{Dichronism3}

\begin{thebibliography}{49}
\bibitem [{Exp()}]{Expansion}%
  \BibitemOpen
  \href@noop {} {}\bibinfo {note} {Formally, all orders of $a/\lambda$ are
  present in the expansion and we have only kept the leading orders in Eq.
  (\ref{const}). For the static response, though, the expansion is
  exact.}\BibitemShut {Stop}%
\bibitem [{chi()}]{chi}%
  \BibitemOpen
  \href@noop {} {}\bibinfo {note} {We denote the chirality by the greek letter
  $\chi$ (from the ancient greek word for hand -
  $\chi\epsilon\acute{\iota}\rho$) which shall not be confused with the
  current-current response matrix $\bm \chi$ or $\chi_0^n$, $\chi_1$,
  $\chi_{xy}$ in Eq. (13).}\BibitemShut {Stop}%
\end{thebibliography}
\end{document}